\documentclass[fleqn,10pt]{wlscirep}
\usepackage[left]{lineno}  
\usepackage{xcolor}
\usepackage{tcolorbox}
\usepackage[english]{babel}
\usepackage[utf8]{inputenc}
\usepackage[T1]{fontenc}
\usepackage{amsmath,amssymb}
\usepackage{bm}
\usepackage{physics}

\title{A lesson from a  small particle about  quantum theory  with strong implications for cosmology}

\author[1,*]{Daniel Sudarsky}
\author[2]{Octavio Guerrero}

\affil[1]{Instituto de Ciencias Nucleares, UNAM POB 70-543, Mexico City 04510, Mexico}
\affil[1]{Departament de Física Quantica i Astrofísica, Universidad de Barcelona, Barcelona 08028, España}
\affil[2]{Department of Physics, The University of Texas at Austin, Austin, TX 78712, USA}

\begin{document}

\begin{abstract}

The establishment of extremely strong bounds on the magnitude of the electric dipole moment of the neutron, a quantity that is of great importance for determining the level of time reversal symmetry respected by the strong interactions, offers an important lesson regarding the manner in which quantum uncertainties are interpreted in the inflationary cosmological account of the generation of the primordial inhomogeneities that give rise to the universe's structure.  The identification of quantum uncertainties with actual stochastic fluctuations, a standard aspect of the current physical account for the emergence of the cosmic structure, is called into question.  This opens the door for novel aspects of physics that are needed in order to provide a satisfactory account that is both conceptually clear and does not conflict with the use of quantum theory in other settings. 

\end{abstract}

\flushbottom
\maketitle

\thispagestyle{empty}

\noindent \textbf{Key points:} 

\begin{itemize}
\item The quantum uncertainties in the electric dipole moment of the neutron (nEDM) are many orders of magnitude larger than the best bounds on the size of that quantity. 
\item This is addressed by understanding the actual experiments as providing \textbf{ a weak measurement} of the corresponding expectation value. 
\item The lesson extracted illustrates the fact that the inflationary account of the generation of the primordial cosmic inhomogeneities is flawed and that novel physics is required to satisfactorily address the issue. 
\end{itemize}

\noindent \textbf{Website summary: The experimental setting of very stringent bounds on the size of the electric dipole moment of the neutron is used to illustrate the problematics of conflating quantum uncertainties with actual stochastic fluctuations.  This, in turn,  is used to argue against the way in which 
quantum uncertainties are treated in the inflationary cosmological account of the generation of the primordial inhomogeneities that give rise to the universe's structure.} 

As everyone who has encountered quantum mechanics can attest, the theory remains a source of puzzlement and heated debates, despite being, without a doubt, one of the most successful and precisely tested theories in the history of science.

 Practicing scientists who work with quantum mechanics frequently rely on something rather vague, known as the ``textbook" or standard interpretation of quantum theory (often mistakenly referred to as the ``Copenhagen interpretation") \cite{SI, peres, wein_cop}.  According to this perspective, the variables of a system acquire definite values only upon measurement; in the meantime, most of the system’s variables simply do not have definite values, evolving,  nonetheless, according to a deterministic law known as the Schrödinger equation. The system is,  in general, characterized quantum mechanically, not by the value of such variables, but by an abstract object known as the ``quantum state"  usually denoted as $|\Psi \rangle $ .  Scientists often construct a variety of mental pictures to gain intuition about the theory and its relation to the world we perceive, i.e. “the world out there”, but the theory itself offers no clear picture of an underlying reality. As many researchers in the foundations of quantum theory, physicists,  and philosophers alike would say, the theory lacks a clear {\it ontology}.

 A particularly thorny issue is how to interpret a variable that does not have a definite value (for instance, the location of a particle that is in a state with indefinite position).  In those situations, the variable in question, \textit{Q}, can be characterized,  at least partially,  by two numbers: one called the associated expectation value  (EV) in the given state, denoted by $\langle  Q \rangle_{\Psi }= \langle \Psi  |Q | \Psi \rangle $ and another called its uncertainty $\Delta Q_\Psi $, often referred to as the ``quantum fluctuation" of that variable.

When the uncertainty in the variable \textit{Q} vanishes, the situation corresponds to one in which the variable \textit{Q} does have a well-defined value (which coincides with $\langle Q \rangle_\Psi $). A  state of a system for which this happens is a rather special one called an  {\it ``eigenstate "} of  $ Q$, and the corresponding associated value $ q$ is called the {\it ``eigenvalue"}. 
However, for generic situations, the uncertainty in most variables does not vanish. In fact, the famous {\it Heisenberg uncertainty principle} shows that the product of uncertainties-- of the so-called conjugate variables (such as position and momentum) -- can never fall below a certain bound. Often, the conditions under which an uncertainty vanishes are considered non-physical.  For example, a particle with a 100\% definite position would, according to the theory, have an infinite expectation value for its energy. This implies a nonzero probability of measuring arbitrarily large energy values, which represents a clear unphysical scenario. 

The basic operational rule of the theory --which, as noted, works remarkably well-- is the following:  if one prepares a large number of copies of the system in the same quantum state, and then proceeds to measure the variable \textit{Q} on each copy,  the outcomes would tend to average to the value given by the expectation value $\langle Q \rangle_\Psi$, with a statistical dispersion of the results that would be given by its uncertainty $\Delta Q_\Psi$. This statistical spread is why the quantum uncertainties are often referred to as ``quantum fluctuations’’.  So far, this seems straightforward. However, one aspect that remains deeply unclear is how to interpret the variable before the system is subjected to a measurement. 

Some people tend to imagine that the variable in question is somehow “randomly jumping” around the expectation value, with instantaneous values deviating by amounts of about the order of magnitude of the uncertainty.  It is often imagined, then, that upon a measurement, one simply  ``catches" one of those random values. Alternatively, others adopt a view where the quantum description is considered to apply just to large collections of identical systems,  as the one described above, while assuming, often implicitly, that each of the elements of the system does have a definite value of \textit{Q}, with the collection of values distributed around the expectation value, with random deviations, again on the scale of $\Delta Q_\Psi$, so that the average of the values coincides with the expectation value \footnote{Sometimes a related point view is expressed by indicating that quantum mechanics applies \textit{ ``in average”} and not to individual systems,  but that seems to ignore the fact that in order to define an average, in the usual sense of the word, each of the elements in the corresponding collection, must have a well-defined value of the quantity in question.  The average grade of a class is not a quantity with a well-defined value until  all students have been assigned their corresponding grade.}. The fact is that, in most cases, the way one chooses to think about these questions makes little or no difference.
 Consequently, many practicing physicists regard foundational issues of this kind as merely philosophical curiosities and peripheral to their professional focus. However, as we will discuss, there are indeed some situations in which a much more delicate perspective is required, and where a nuance in the analysis is paramount. For example, subtle foundational perspectives are indispensable when analyzing the implications of Bell inequality experiments \cite{Maudlin}, the GHZ gedankenexperiment\cite{GHZ}, or to the Kochen-Specker Theorem\cite{KS}. 

One of the most important fields in which the rather naive kinds of views described above are almost universally adopted is cosmology.  More precisely, this occurs when dealing with the so-called inflationary epoch, an early stage of enormous and accelerated expansion in our Universe's history. 
Inflation is thought to naturally set the stage for the subsequent evolution, known as the old Big Bang cosmological model, by diluting, to an extreme degree, pre-existing features (such as very heavy objects known as topological defects), as well as all deviations from homogeneity and isotropy, and making the spatial geometry exponentially close to a flat one\footnote{By spatial geometry one  refers to that characterizing the   " surfaces" of   constant  `` cosmic  time", the  time that  would  be   measured  by co-moving observers ( i.e. those for which the universe looks  maximally isotropic). One can think of  one such  surface   defined by,  say,  the  value of the  temperature of CMB radiation  at its points.}.  This accelerated expansion is meant to be driven by the potential energy of the so-called \textit{inflaton} scalar field –a field analogous to the electromagnetic field, but lacking directionality, and thus, in a sense, simpler.  The inflaton is supposed to be ``excited’’ only in its homogeneous and isotropic configuration (or mode), known as the zero mode, while the remaining degrees of freedom are taken to be in the so-called vacuum state (as are all other matter degrees of freedom). This state is expected to result from the extremely rapid and accelerated expansion of the Universe, characterized by a growth in scale by a factor of about $e^{60}$, within a small fraction of a second, at very early times in cosmic history. 
The matter  content of that stage of the universe is then described by what is called an adiabatic vacuum state  of all matter  including the  very field that  is  responsible for the accelerated expansion, the inflaton.   The  enormously accelerated expansion process causes the above-mentioned dilution, leading, therefore, to the essentially complete homogenization of the universe (and at the same time making its spatial curvature become insignificantly small, and thus taken, in practice, to vanish exactly).
In short, this is a characterization of the so-called inflationary cosmic epoch, now considered as an essential component of the universe’s history \cite{guth, linde, LythLiddle2009} \footnote{The fact that this homogenization is characterized by such a large number allows it, in practice, to be treated as absolute, since any deviation from perfect homogeneity left after the inflationary process would be insignificant in comparison with the very important departures of homogeneity connected to the emergence of cosmic structure,  that we will be discussing next.}.

The above schematic account of the inflationary epoch, as accepted in our current most widely cosmological model, immediately gives rise to a fundamental question: If the universe was completely homogeneous and isotropic, both in its classical description (the background geometry) and in its quantum description (the vacuum state of all quantum fields,  with the sole exception of the highly excited "zero-mode "  --or homogeneous and isotropic mode--  of the inflaton) then how did the richly structured Universe we observe arise? Where do we find room to account for the plethora of structures in the cosmos? These include galaxy clusters, galaxies, gas clouds, stars and planets (including our solar system), and even ourselves, the human being.  In fact, we can now  observe small traces of the corresponding early imprints  of such structure,  which are  imprinted in the Cosmic Microwave Background  (CMB) \footnote{That is, the radiation emitted when the universe was about  1100  times smaller than it is today (and characterized by a temperature of about $3000 K^o$).  A moment in which electrons became bounded to nuclei to form the first atoms, and which is currently the subject of intense studies by dedicated instruments like the Planck satellite\cite{planck} and others.}.

Well, inflationary cosmology offers a proposed answer to the question:  the primordial seeds of all the structure we observe trace their origin to the quantum fluctuations associated with the state of the inflaton field (and with other fields, including spacetime metric perturbations, that are treated alongside).  The key idea is that one can compute such quantum uncertainties and reinterpret them as some sort of classical stochastic fluctuations (i.e. as representing some kind of variations of the physical conditions in space and time) corresponding to the involved fields, and one uses the resulting values to estimate the characteristic spectrum (i.e., the amplitude of variations decomposed into plane waves of specific wavelength and direction). The results show remarkable agreement with both the observations of the large scale structure of the universe \cite{BAO} and the studies of the early imprint of such features in the CMB\cite{planck}. This agreement is often regarded as one of the most astonishing successes of inflationary cosmology.

However, a moment’s reflection reveals a conceptual gap at the heart of this account. Namely, what exactly justifies the identification of the quantum uncertainties with classical stochastic fluctuations?  

Given that the quantum state of the system is completely homogeneous and isotropic, what justifies extracting or reading from it any meaningful measure of inhomogeneity and isotropy? 

Instead of engaging here in a detailed discussion of the preceding question, we next turn to what may seem like a completely unrelated issue, one that is, at first glance, also puzzling, and offers a dramatic lesson  and  a warning  regarding an incorrect  way  to think about quantum theory.  We refer to one of the fundamental quantities associated with one of the most common particles in the universe: The electric dipole moment (EDM) of the neutron.  This quantity has attracted great attention for decades, as it relates to a fundamental symmetry that was, for a long time, thought to be an exact symmetry of the strong interactions, but with the advent of QCD, some good reasons emerged to doubt if that was the case\cite{CP-strong}.  We are referring to the symmetry known as CP symmetry, a combination of charge conjugation C, corresponding to exchanging particles with antiparticles, followed by parity P, which inverts spatial coordinates (left and right). For some time, it was thought that CP was an exact symmetry of all physical interactions, until it was discovered in 1964 that weak interactions violate CP slightly, as evidenced in experiments involving the neutral Kaon system. However, this violation is so minuscule that it plays no significant role in our present discussion.

The electric dipole moment of the neutron EDM, denoted $d_n$, is considered one of the most promising testing grounds for detecting a violation of CP in the strong  (or even electromagnetic) interactions. This has motivated the quest for detecting any evidence of a nonzero value of the EDM of the neutron. More precisely, there is a simple theorem indicating that if CP is an exact symmetry of nature, then the expectation value of the neutron EDM must vanish identically. Since the only known CP violation in nature lies within the weak interaction—connected to heavier neutral mesons—this implies that any contribution to the neutron EDM from the weak force must be exceedingly small   (estimated at $10^{-32}  e\cdot cm$), which is completely negligible for our purposes.

 Remarkably, nowadays our experimental colleagues have been able to set extremely stringent bounds on $d_n$ reaching an amazing accuracy, with the latest report establishing $ d_n < 10^{-26}  e\cdot cm$ (where $e$ is the magnitude of the electron's charge). 

 Yet, when considering the situation within a quantum mechanical context, a conceptual issue arises. At the quantum level, one must also consider the uncertainty in the value of the Electric Dipole Moment of the Neutron. A simple estimate considering the neutron’s size times their constituent valence quarks charges (two up quarks with charge $ (1/3) e $ and one down quark $(-2/3)e $) indicates that this uncertainty $ \Delta_{d_n}$  should be of order $10^{-15}    e\cdot cm$. That is, the quantum uncertainty is at least TEN orders of magnitude larger than the experimental upper bound on $ d_n$. This raises the puzzle: even if the expectation value of the neutron EDM was exactly  $0$,  wouldn’t quantum mechanics prevent us from setting a bound any tighter than about  $10^{-15}   e\cdot cm $?. How, then, are our experimental colleagues able to set bounds ten orders of magnitude more stringent?

 At this point, we should first clarify a basic concept. Even if a particle is prepared in an initial state in which, say,  its position has an uncertainty  $ \Delta Q$  with expectation value  $\langle Q \rangle_\Psi = Q_0$, it is still entirely possible to measure its position with a much higher precision, say  $ \delta Q << \Delta Q$.  However, the result of such measurement is quite likely to deviate from $Q_0$  by an amount of order $\Delta Q $.  If we repeat the experiment across a large collection of identically prepared particles, each individual measurement may be highly precise (within $\delta Q$), yet the distribution of outcomes will reflect the broader spread $\Delta Q$ centered on $Q_0$.

Then, if that is what our experimental colleagues encountered, they would have to report that their experiment yielded a bound on $d_n$ (assuming its expectation value vanishes or is set by the tiny weak interaction effects), which is of order $10^{-15} e\,cm$.
Then, there seems to be a good reason to be a bit puzzled. Are we witnessing a breakdown of quantum theory?  Are we grossly underestimating the quantum uncertainty?  We do not think so.  Is our "back of the envelope" estimate wrong by more than ten orders of magnitude? 
As noted in \cite{us}, a statistical dispersion of results of the corresponding  magnitude  would have  impeded the  achievement of  such tight experimental  bounds, and moreover, forcing the system into  becoming  an almost exact  eigenstate  of the EDM  operator  would require a  dramatic change in the  hadron internal structure,   so  that it  could no longer be called a neutron. Thus, accepting that result, we must wonder what is going on?.


However, before continuing, let us dig deeper into what the experimentalists do and what they observe:

\begin{tcolorbox}[title=The nEDM experiment, colback=brown!10, colframe=brown!50!black]
 The basic  experimental principle is that  the neutron has a permanent electric dipole moment \(d_n\neq 0\), its spin would precess a little faster or slower depending on
whether the electric field \(\mathbf E\) points with or against the magnetic field \(\mathbf B\).
The Larmor frequencies for  the two configurations are

$$
f_{\uparrow\uparrow,\uparrow\downarrow}
= \frac{1}{ \pi}|\mu_n \mathbf{B}\, \pm\, d_n \mathbf{E}|
$$
so flipping \(\mathbf E\) flips the EDM term while the magnetic term stays the same.

So far, the most precise experimental value for the nEDM \cite{nedmpsi} was found at the Paul Sherrer Institut (leading to  $ d_n=-0.21 \pm 1.82 \times 10^{-26}  e.cm $).  For a simplified description of the experiment, we start with a certain population of (``ultracold" i.e. very slow moving ) neutrons stored between high-voltage electrodes in a very uniform magnetic field. The spin direction of these neutrons is aligned  let's say  in $+\hat{z}$ direction.
Then, a radio-frequency pulse tips the spin direction into the plane where they precess, let's say along the $x-y$ plane. Once this step has finished, they let the neutrons precess freely (about $\hat{z}$) for a certain period of time in which each neutron accumulates a phase $\phi$ proportional to the nEDM. 
After the precession stage, a second pulse converts the accumulated phase into a population
difference that is countable as “spin-up” and “spin-down” neutrons. At this point, the chamber is emptied and the neutrons are guided into two arms, each containing a magnetized foil that transmits one spin state and reflects the other, allowing only the desired state to pass through and be counted by the detector. The reflected neutrons are transported from one arm to the other. This experimental arrangement allows for simultaneous counting, minimizing any bias due to variations in the neutron flux\cite{device}.  This count is used to define the \textit{asymmetry}

$$A = \frac{N_\uparrow - N_\downarrow}{N_{\uparrow} + N_{\downarrow}} \simeq \cos{\phi}.$$

On the other hand, a small amount of \(^{199}\)Hg gas is placed in  the same volume and acts as a built-in magnetic clock; its precession frequency \(f_{\rm Hg}\) is
continuously  measured.

Comparing the neutron frequency to that of  the mercury cancels most magnetic field  variations. The analysis uses
$$
R \;\equiv\; \frac{f_n}{f_{\rm Hg}}
\;\approx\;
\frac{|\gamma_n|}{|\gamma_{\rm Hg}|}
\;\pm\; \frac{d_n\,E}{\pi\,f_{\rm Hg}}
\;+\; \Delta R_{\rm sys},
$$
where \(\Delta R_{\rm sys}\) denotes small, well-characterized corrections (such as those related to tiny field variations).
Alternating the direction of \(\mathbf E\) and fitting the change in \(R\) isolates the term proportional to \(d_n\)\cite{abel_nedm_2019}.

\end{tcolorbox}

In \cite{us}, the issue has been analyzed carefully, and a clear resolution to this seemingly paradoxical situation was found.  It turns out to lie on a subtle aspect of the discussion, often overlooked in both theoretical discussions and experimental analyses.  Specifically, it involves the distinction between the "weak value'' of a quantity,  coinciding in this case with the expectation value $ \langle \hat {\vec{d}}  \rangle$ of the neutron EDM in its ground state, and the {\bf ``actual'' value} of the EDM of the neutron resulting from a standard measurement. That is, the difference between the (somewhat vague) number assigned by quantum theory to a quantity that does not possess a definite value before measurement, and the value that might be obtained when the system is subjected to a measurement, which, in doing so, forces the system to acquire a definite value which is reflected in the outcome. This is the standard way in which measurements enter quantum theory, and for the sake of the remainder of the discussion, we will refer to these as "strong measurements".  

At this point, we must turn our attention to a quite different notion of measurement introduced in
\cite{A-V}, which exhibits rather distinct properties, and which leads to a rather different set of considerations in comparison with the more standard strong measurements.  In a weak measurement, the system of interest $S $ interacts only very weakly with an auxiliary system $A$  (the ``ancilla’’), and information about the state of the system $S$ is extracted by performing strong measurements on $A$.  In particular, in the simplest versions of a weak measurement one extracts directly the expectation value of the quantity of interest in the initial state of the system $S$.   Significantly, this procedure can be designed so that the state of $S$ is as mildly disturbed as desired. 
     
In short, a weak measurement is fundamentally a different kind of measurement of a certain quantum variable.  Since it is highly indirect, it minimally perturbs the system's state,  and yields, essentially, the expectation value of the variable in the undisturbed state of the system. In contrast, the strong measurement  of   an  attribute $Q$  of  the system S,  forces the system to ``collapse’’ into an eigenstate of  $Q$ , yielding a definite value and changing the state of the system. The result of a weak measurement is sometimes called the ``weak value" of the corresponding quantity, and quantum theory imposes no limitation on the accuracy with which said quantity might be measured.

In \cite{us}, it was argued that the experimental results previously discussed refer, in fact, to the weak value of the EDM of the neutron, not its actual (strong measurement) value. Indeed, what the theorem we mentioned above actually says is that if CP is a good symmetry of nature, then it is the expectation  value of the EDM of the neutron that needs to vanish.



\begin{tcolorbox}[title=Quantum analysis of   a simplified version of  the experiment  , colback=brown!10, colframe=brown!50!black]

As a first simplification, we will focus only on the neutron, its internal degrees of freedom, and the spatial orientation of its spin. The external electromagnetic field  is treated classically, taking  a vanishing   magnetic field so that the precession of the spin depends  just on {$\vec{D}$} and {$\vec{E}$}.
Next,  we chose to   distinguish between the neutron's internal degrees of freedom that determine the magnitude of the EDM and the degrees of freedom that characterize its  spatial orientation ( one playing the role of the quantity  of  interest { $ O$}, and the other  of the  auxiliary one{ $ A$}).

This approach relies the Wigner-Eckart theorem,  which  allows us to express
the expectation value  of the  electric  dipole moment as

$$
\langle \hat{\vec {D_i}} \rangle =  \frac{\langle\hat {\cal D} \rangle }{j(j+1)}\langle\hat {\vec {J_i}}\rangle
$$
  in terms of  the scalar  operator  
$\hat{\cal{D}}\equiv(\hat{\vec{S}} \cdot \hat{\vec{D}}+\hat{\vec{D}} \cdot \hat{\vec{S}})/2$.
and 
where{ $\hat{\vec{J}}$} represents the angular momentum operator, which, in our context, corresponds to the neutron's spin ($\hat{\vec{S}}$), thus{$j=1/2$}.

Next, we   exhibit how the quantum uncertainty of the EDM  enters into the analysis of the weak measurement illustrating its  influence on the accuracy with which the quantity of interest can be measured.
We write the quantum state of the neutron, explicitly separating the degrees of freedom as indicated above. That is, we express the state of a neutron  at rest in the lab and with a spin orientation { $\vec{\sigma}$} as:
$$
|\psi\rangle =   N \int d(d)   e^{-(d- d_n)^2/ 2\Delta^2}  |d\rangle \otimes | \vec \sigma\rangle  , 
$$

here,{$|d\rangle$} represents the eigenstates of{$\hat{\cal{D}}$} (normalized according to{$\langle d' | d \rangle = \delta(d - d')$}), {$ d_n= \langle n | \hat{\cal{D}} | n \rangle$} is the expectation value of the dipole moment scalar in the neutron's  ground  state, and {$\Delta$} represents its quantum uncertainty ( { $N$} a normalization factor ).
Here$|\vec{\sigma}\rangle$ denotes eigenstates of the spin along the direction {$\vec{\sigma}$}, 

The  initial  state representing a  neutron  (prepared in the lab. to  be  essentially  at rest and with  its spin pointing in the direction {$\vec{\sigma} = +z$}).  
An  external electric field in the direction {$ +y $}   is applied during the time interval 
{$ [0, t_f ] $},  ( switched on and off slowly enough to avoid transient effects).
          The interaction of the  neutron with  this  electric field ( taken to point in the direction $+y$)   is  represented by the  Hamiltonian: ${\hat{ H_i }=  \vec E (t)\cdot\hat{\vec D}  }$

We are interested in computing  the probability amplitude  for ``finding" the  neutron in  the   state with 
{$\vec{\sigma} = -z$}),  
at {$t= t_f$}. The relevant {\it transition}  amplitude is :
$ {\cal{ A} } = \langle \psi'| e^{i \int  H_i  dt}  |\psi_0\rangle $
 which can be  explicitly  evaluated   yielding a corresponding probability
$$
{\cal{ P} } 
= {\sin{( {d_n} \xi )}}^2 e^{-\xi^2 {\Delta}^2}.
$$
where  {$ \xi  \equiv \frac{1}{2 \sqrt{1/2 (1/2+1) }}  \int_0^{t_f}   E(t)  dt $}.
Note that   $ {\cal{ P} } $  vanishes identically if  CP  symmetry is present  (and $d_n =0$),  quantum  fluctuations  not withstanding.  
 
The result explains the feasibility of measuring  $ d_n$  to arbitrarily precision, regardless of the magnitude of $\Delta$.  The  transition probability can be  measured with arbitrarily high accuracy,  ( using a sufficiently large number of neutrons).
The  amplitude depends  on {$\Delta$}, which exact  value is unknown, but {$\xi$} depends  on the applied electric field's magnitude and the duration of its application, and the functional  dependence is   different for  the two parameters,{$d_n$} and {$\Delta$}. 

\end{tcolorbox}

Next, let us recall the so-called ``statistical interpretation'' of quantum theory, or more generally, the notion that, when a system is in a state exhibiting quantum fluctuations of a certain variable, that variable may be regarded as taking random values, either within an ensemble of identical systems, or for a single system, at different times. This picture clearly cannot describe the neutron-EDM experiment we have just described. If it were the case, we would have seen quite a large quantity of spin-flipped neutrons from the large ensemble of neutrons that would sample EDM values ranging from  $0$ up to order $10^{-16} e\cdot cm$. The point, however, is that the actual experimental results indicate that, in the simplified version we have analyzed,   besides some small level of experimental noise, not a single spin-flipped neutron that could be blamed   to a non-zero value of the EDM  would be observed.
 
This shows, in quite  an unmistakable fashion, that such views regarding quantum theory are simply untenable.  That itself is a very valuable lesson on the foundations of quantum theory extracted from a remarkable technological achievement and exquisite precision of our experimental colleagues working on the neutron-EDM programme (and similar systems).

However, the lesson reaches farther. The fact of the matter is that in contemporary cosmology, the entire account for the randomness that seeds the large-scale distribution of matter relies on, precisely, this kind of statistical interpretation of quantum fluctuations.

As we noted in the initial discussion,  according to the accepted ``lore in cosmology” the seeds of all cosmic structure are supposed to emerge out of the quantum fluctuations in the  ``adiabatic vacuum state of the fields”.  The problem is that,  the evolution provided by the Schrodinger equation for a  state that was, at one time,
homogeneous and isotropic will preserve those features of the state for all future time. Most cosmologist, while accepting this fact, would argue, in turn,  that one  must consider the quantum fluctuations, and  that these serve to somehow address the issue \cite{kiefer-polarski}. However when considering that, we have to note that quantum fluctuations might become actual stochastic variations within an ensemble \footnote{Considering for  the argument’s  sake  that the universe, at that  stage,  might be regarded  as  just an ensemble of small regions.},  only as  a result of  a measurement (see detailed  discussions in \cite{PSS,Shortcomings})\footnote{Cosmologist often invoque other aspects of the problem, like decoherence or squeezing of the quantum  state, but a careful examination shows  (see \cite{Shortcomings, debunking} ) that such considerations  do not really  serve to address the issue. }.   On the other hand,  during the early epochs in the universe we are referring to,  there were no observers to carry out measurements, nor were there anything like measuring devices!. Could we nonetheless adopt the view that the presence of quantum fluctuations somehow indicates that the fields had definite values in  each  such region? Or that the values of the fields at such points were simply varying wildly within a range characterized by the corresponding  quantum uncertainties?   The  answer  is \textit{NO}, that is, in fact, the lesson we learned by considering the quantum uncertainty of the  Neutron’s Electric  Dipole Moment.   

With that lesson in mind, let us revisit the inflationary account of the “origin of the primordial seeds of cosmic structure”.  It is now clear that the central step we highlighted in that account cannot be justified as part of a generic understanding of quantum theory. We have seen that one cannot simply adopt the view that, in the absence of measurements,  the quantum uncertainties could be taken to represent stochastic fluctuations. Such a posture would lead us astray when considering situations such as those relevant for experiments like those providing bounds on the  Neutron’s EDM.  However, there remains an indisputable fact that the actual inflationary  ``predictions'' regarding the primordial density fluctuations that seed the growth of cosmic structure are, even if not satisfactorily justified,   remarkably successful in accounting both for the large scale distribution of matter in the universe\cite{BAO-DESI}  and their earliest imprints observable in the Cosmic  Microwave Background radiation \cite{COBE, WMAP},  so we must look for novel aspects of physics to account for those facts in a conceptually clear and satisfactory manner, while ensuring that we do not rely on interpretational steeps that would lead to conflicts with other established empirical results (such as those we have discussed here in relation with the neutron EDM).   The recognition of a shortcoming in our best understanding is an important step in the search for a more profound understanding of both the fundamental laws of nature 
and the manner in which they account for the history of our universe \footnote {In this regard, we can mention recent works based on the  Bohmian Mechanics version of quantum theory \cite{pinto-neto} and others, focusing on spontaneous collapse theories \cite{Martin, okon_benefits_2014, singh_quantum_2018}.  At this point, it is not completely clear that those accounts are fully satisfactory, and further research along those lines is needed to ascertain their viability.}

That is a very profound lesson, and it forces us to recognize a serious flaw in our best account of the evolution of our universe.  For that,  we are indebted to the tiny neutron (and the scientists who have made such heroic efforts in carrying out the empirically powerful results about its EDM). In facing that issue, we must then recognize that we need,  on the one hand, to refine our account of quantum theory, and, at the same time, we must find a theoretical setting which would accommodate these two sets of results (the absence of stochastic behavior of the neutron's EDM, despite the quantum uncertainties and the actual breakdown of homogeneity and isotropy in the state of the 
inflationary field in the early universe,  in accordance with the observations of large scale structure and of the CMB). The times when we could regard the conceptual difficulties that have long been known to plague the theory as issues which have ``only philosophical interest", and as being completely irrelevant for the practical use of the theory,  are simply gone. These questions, which have drawn the attention of philosophers of physics and physicists concerned with foundational aspects of quantum theory, should be regarded as serious conceptual challenges, particularly when one aims to describe a system’s behavior before any measurement can be said to have taken place. A general and consistent view, applicable across these diverse deployments of quantum theory, must be sought. Beyond the case of the neutron’s EDM, such contexts include the inflationary cosmological setting and others not explicitly discussed here, such as black hole physics (see \cite{York, Thompson}) and broader formalisms like stochastic gravity \cite{hu_stochastic_2004}, in which quantum fluctuations play a central role.
 
One might view this situation  (where a serious shortcoming in our current understanding is recognized) with certain anguish,  but we prefer to look at its bright side.  We are confronted here with a situation involving gravitation,  quantum theory,   and empirical data \footnote{We thank Jérôme  Martin for this observation.},  that might turn out to  offer  important  lessons  about   deeper levels of  the workings of nature, and  perhaps   clues  about the long sought-for theory capable of a unified  treatment of  both pillars of modern physics:   General Relativity and the Quantum.  Desperate for  any  {\it "hints  of   novel  aspects of physics"} the physics community   at large might  be overlooking a very important  opportunity.   The {\it big lesson}  from a small particle  about the largest of all  systems  ever studied  by physics,  should be appreciated  and  considered  with due care.

\bibliography{sample}

\end{document}